\documentclass[11pt]{article}

\usepackage{graphicx}
\usepackage{amsmath}
\usepackage{amsfonts}

\begin{document}

\title{An application of Malliavin Calculus to Finance
}
\author{Arturo Kohatsu-Higa${}^{\dagger}$ and Miquel Montero${}^{\ddag,\dagger}$\\
%EndAName
${}^{\dagger}$Department of Economics, Universitat Pompeu Fabra, \\
Ram\'{o}n Trias Fargas 25-27, 08005 Barcelona, Spain\\
${}^{\ddag}$Departament de F\'{\i}sica Fonamental, Universitat de Barcelona, \\
Diagonal 647, 08028 Barcelona, Spain
}
\date{\empty}
\maketitle

\begin{abstract}
In this article, we give a brief informal introduction to
Malliavin Calculus for newcomers. We apply these ideas to the
simulation of Greeks in Finance. First to European-type options
where formulas can be computed explicitly and therefore can serve
as testing ground. Later we study the case of Asian options where
close formulas are not available. The Greeks are computed through
Monte Carlo simulation.
\end{abstract}

\section{Introduction}

Malliavin Calculus is an area of research which for many years has
been considered highly theoretical and technical from the
mathematical point of view. In recent years it has become clear
that there are various applications of Malliavin Calculus as far
as the integration by parts formula is concerned. Nevertheless it
is still considered by the general mathematical audience and
practitioners a field where is hard to grasp the basic ideas or to
obtain new contributions.

In this article we present an informal introduction to Malliavin
Calculus which we hope can open the area to practitioners. For
serious mathematical approaches to the topic we refer the readers
to the authoritative books on the matter, e.g. \cite{Ikeda, Nualart, Malliavin}. We have
tried to take the spirit of the issues to try to explain in simple
terms the elements of the theory.

\section{Malliavin calculus}
The most common concise way Malliavin Calculus is presented in a
research paper is as follows. Let $W=\{W_{t}\}_{t\in \lbrack
0,1]}$ be a standard one-dimensional Brownian motion defined on a
complete probability space $(\Omega ,\mathcal{F},P)$. Assume
$\mathcal{F}=\{\mathcal{F}_{t}\}_{t\in \lbrack 0,1]}$ is generated
by
$W$. Let $\mathcal{S}$ be the space of random variables of the form $%
F=f(W_{t_{1}},\ldots ,W_{t_{n}})$, where $f$ is smooth. For $F\in \mathcal{S}
$, $D_{t}F=\sum_{i=1}^{n}\frac{\partial }{\partial x_{i}}f(W_{t_{1}},\ldots
,W_{t_{n}}){\bf 1}_{[0,t_{i}]}(t)$. For $k\in \mathbb{Z}_{+}$, $p\geq 1$, let $%
\mathbb{D}^{k,p}$ be the completion of $\mathcal{S}$ with the respect to the
norm
\begin{equation*}
\Vert F\Vert _{k,p}=(E[|F|^{p}]+E[(\sum_{j=1}^{k}\int_{0}^{1}\ldots
\int_{0}^{1}|D_{s_{1},\ldots ,s_{j}}^{j}F|^{2}ds_{1}\ldots
ds_{j})^{p/2}])^{1/p},
\end{equation*}
where $D_{t_{1},\ldots ,t_{j}}^{j}F=D_{t_{1}}\ldots D_{t_{j}}F$. We let $%
\Vert F\Vert _{0,p}=(E[F^{p}])^{1/p}=\Vert F\Vert _{p}$ and $\mathbb{D}%
^{\infty }=\cap _{k,p}\mathbb{D}^{k,p}$. For processes $u=\{u_{t}\}_{t\in
\lbrack 0,1]}$ on $(\Omega ,\mathcal{F},P)$, $\mathbb{D}_{L^{2}([0,1])}^{k,p}
$ is defined as $\mathbb{D}^{k,p}$ but with norm $\Vert u\Vert
_{k,p,L^{2}([0,1])}=(E[\Vert u\Vert
_{L^{2}([0,1])}^{p}]+E[(\sum_{j=1}^{k}\int_{0}^{1}\ldots \int_{0}^{1}\Vert
D_{s_{1},\ldots ,s_{j}}^{j}u\Vert _{L^{2}([0,1])}^{2}ds_{1}\ldots ds_{j})^{p/2}])^{1/p}$.

We denote by $D^* (u)$ the Skorokhod integral or the adjoint operator of $D$.
This adjoint operator behaves like a stochastic integral. In fact, if $u_{t}$ is $\mathcal{F}%
_{t}$ adapted, then $D^* (u)=\int_{0}^{1}u_{t}dW_{t}$, the It\^{o}
integral of $u$; see e.g. \cite{Nualart}. Here we write $D^*
(u)=\int_{0}^{1}u_{t}dW_{t}$, even if $u_{t}$ is not
$\mathcal{F}_{t}$ adapted. There are other anticipating integrals
that have some relationship with this one as e.g. the Ogawa
symmetric integral. Of the formulas we will use, the following are
worth mentioning,
\begin{equation}
\int_{0}^{1}Fu_{t}dW_{t}=F\int_{0}^{1}u_{t}dW_{t}-\int_{0}^{1}(D_{t}F)u_{t}dt,
\label{D2}
\end{equation}
for $F\in \mathbb{D}^{1,2}$ and $E[F^{2}\int_{0}^{1}u_{t}^{2}dt]<\infty $ ---
see e.g. \cite{Nualart}---; and
\begin{equation}
\label{dp} E\left[\int_{0}^{1}(D_{t}F)u_{t}dt\right]=E[F D^* (u)].
\end{equation}

As a byproduct of all the above formulas one obtains the integration by
parts formula. For this, we say that $F$ is smooth if $F\in \mathbb{D}%
^{\infty }$. For a real random variable $F\in \mathbb{D}^{1,2}$, we denote by $%
\psi _{F}$ the Malliavin covariance matrix associated with $F$. That is, $%
\psi _{F}=<DF,DF>_{L^{2}[0,1]\times \mathbb{R}}$. One says that the random
variable is non-degenerate if $F\in $ $\mathbb{D}^{\infty }$ and the matrix $%
\psi _{F}$ is invertible a.s. and $(\det \psi _{F})^{-1}\in \cap _{p\geq
1}L^{p}(\Omega )$.

The integration by parts formula of Malliavin Calculus can be briefly
described as follows. Suppose that $F$ is a non-degenerate random variable
and $G\in $ $\mathbb{D}^{\infty }$. Then for any function $g\in
C_{p}^{\infty }(\mathbb{R}^{q})$ and a finite sequence of multi-indexes $%
\beta $, we have that there exists a random variable $H^{\beta }(F,G)$ so
that
\begin{equation*}
E[g^{\beta }(F)G]=E[g(F)H^{\beta }(F,G)] \text{ with}
\end{equation*}
\begin{equation*}
\left\| H^{\beta }(F,G)\right\| _{n,p}\leq C(n,p,\beta )\left\| \det (\psi
_{F})^{-1}\right\| _{p^{\prime }}^{a^{\prime }}\left\| F\right\|
_{d,b}^{a}\left\| G\right\| _{d^{\prime },b^{\prime }},
\end{equation*}
for some constants $C(n,p,\beta )$, $a$, $b$, $d$, $p^{\prime }$, $a^{\prime
}$, $b^{\prime }$, $d^{\prime }$ and $\beta \in \cup _{n\geq
1}\{1,...,q\}^{n}$. Here $g^{\beta }$ denotes the high order derivative of
order $l(\beta )$, the length of the multi-index $\beta$, and whose partial derivatives are taken according the index vector $\beta $.

A gentler introduction may say that the idea behind the operator
$D$ is to differentiate a random variable with respect to the
underlying noise being this generated by the Wiener process $W$.
Therefore heuristically one may think that $D_s
``="\frac{\partial}{\partial(dW_s)}$. With this in mind one can
guess how to differentiate various random variables. Some examples
are
\begin{eqnarray*}
D_{t}W_{t} &=&1 ,\\
D_{t}f(W_{t}) &=&f^{\prime }(W_{t}), \text{ and}\\
D_{s}\left(
\int_{0}^{1}f(W_{u})dW_{u}\right)&=&\int_{s}^{1}f^{\prime
}(W_{u})dW_{u}+f(W_{s}).
\end{eqnarray*}
Here $f$ is a $C^1_b$ function. A way to understand any
integration by parts formula is through the following general
definition.

{\bf Definition} We will say that given two random variables $X$
and $Y$, the integration by parts formula is valid if for any
smooth function $f$ with bounded derivatives we have that
\begin{equation*}
E[f^{\prime }(X)Y]=E[f(X)H],
\end{equation*}
for some random variable $H\equiv H(X,Y)$.

One can deduce an integration by parts formula through the duality
principle (\ref{dp}). That is, let $Z=f(X)$. Then using the chain
rule we have
\begin{equation*}
D_{s}Z=f^{\prime }(X)D_{s}X.
\end{equation*}
From here we multiply the above by $Yh(s)$ where $h$ is a process
to be chosen appropiately. Then
\begin{equation*}
D_{s}ZYh(s)=f^{\prime }(X)D_{s}XYh(s).
\end{equation*}
Integrating this for $s\in \lbrack 0,1]$, we have that
\begin{equation*}
\int_{0}^{1}D_{s}ZYh(s)ds=\int_{0}^{1}f^{\prime
}(X)D_{s}XYh(s)ds=f^{\prime }(X)Y\int_{0}^{1}h(s)D_{s}Xds, \text{ then}
\end{equation*}
\begin{equation*}
\int_{0}^{1}\frac{YD_{s}Zh(s)}{\int_{0}^{1}h(v)D_{v}Xdv}ds=f^{\prime
}(X)Y, \text{ therefore}
\end{equation*}
\begin{equation*}
E<DZ,u>_{L^{2}[0,1]}=E\left[ f^{\prime }(X)Y\right],
\end{equation*}
with
\begin{equation*}
u_{s}=\frac{Yh(s)}{\int_{0}^{1}h(v)D_{v}Xdv}.
\end{equation*}
Finally, we have that if $D^{\ast }$ is the adjoint operator of
$D$ ---see equation (\ref{dp})---, then
\begin{equation*}
E[ZD^{\ast }(u)]=E\left[ f^{\prime }(X)Y\right], \text{ and}
\end{equation*}
\begin{equation*}
E\left[ f(X)D^{\ast }\left(
\frac{Y h(\cdot)}{\int_{0}^{1}h(v)D_{v}Xdv}\right) \right] =E\left[
f^{\prime }(X)Y\right].
\end{equation*}
This also means that in particular for $h\equiv 1$ we have that
\begin{equation}
\label{D1} H\equiv H(X,Y)=D^{\ast }\left(
\frac{Y}{\int_{0}^{1}D_{v}Xdv}\right).
\end{equation}
If one has higher order derivatives then one has to repeat this
procedure iteratively. The use of the norms in the spaces
$\mathbb{D}^{n,p}$ is necessary in order to prove that the above
expectations are finite (in particular the ones related to $H$).
Note that the integral $\int_{0}^{1}h(v)D_{v}Xdv$ should not be
degenerate with probability one. Otherwise the above argument is
bound to fail. The process $h$ that appears in this calculation
is a parameter process that can be chosen so as to obtain this
non-degeneracy. In the particular case that $h(v)=D_{v}X$ one
obtains the so called Malliavin covariance matrix.

In conclusion one can build different integrations by parts
formulas depending on how we choose this process $h$. In the next
section we use this formula in order to apply it to a concrete
problem in Finance.
\section{Greeks in Finance}
European options are contracts that are signed between two parties
(usually a bank and a customer) that allows to obtain certain
monetary benefits if the price of certain asset fall above (call
option) or below (put option) a certain fixed value, the
strike price, at a certain fixed date, the expiration time. A
Greek is the derivative of an option price with respect to a
parameter. In general, let $X\equiv X(\alpha )$ be a random
variable that depends on a parameter $\alpha .$ Suppose that the
option price is computed through a payoff function in the
following form ${\cal P} (\alpha )=E[\Phi(X(\alpha),\alpha)]$ where
$\Phi$ is generally non-smooth. A Greek is therefore  a measure of the
sensibility of this price with respect to its parameters. In
particular, it could serve to prevent future dangers in the position of a
company holding these options. The problem of computing Greeks in Finance has been studied by various authors: \cite{Broadie, Benhamou, LEcuyer, Fournie99, Fournie01}, among others. Let us take a clooser look at the problem. If the Leibnitz rule of interchange
between expectation and differentiation were true then we would
have
\begin{equation}
\frac{\partial {\cal P} (\alpha )}{\partial \alpha }=\frac{\partial
E[\Phi (X(\alpha ),\alpha )]}{\partial \alpha }=E\left[\Phi'(X(\alpha
),\alpha )\frac{\partial X (\alpha )}{\partial \alpha
}+\frac{\partial \Phi (X (\alpha ),\alpha)}{\partial \alpha }\right].\label{Direct}
\end{equation}
When the above expression does not have a close formula then one may
start thinking in performing some Monte Carlo simulations in order
to approximate the above quantity. If $\Phi$ is somewhat regular
then we can use the last expression above to do this: we shall call this procedure the ``direct method".
Unfortunately in various cases $\Phi$ is not differentiable. Then
one can resort to the use of the middle expression above to
generate what is known as finite difference method. This method
has been somewhat successful in the recent past but we would like
to discuss here the application of the integration by parts pesented before in
order to compute these derivatives.

\section{The European-style options}
We shall illustrate how this procedure works by choosing a very
special subset of the large family of Financial derivatives: what
we have called {\it European-style\/} options. In this class of
derivatives we will find all the options whose payoff function
depends only on the value of the underlying at the expiration time
$T$, which is previously fixed. Examples of the European-type options are the {\it
vanillas\/} ---the more classical European calls and puts---, or
the {\it binaries\/} ---the so called ``cash-or-nothing"
options---, among others. These options will differ, for instance, from
the {\it American-style\/} options, where the execution time is
not fixed but belongs to an interval; and also from the {\it
Asian-style\/} options where the payoff depends on some average
of the value of the asset in a given period of time. We will
return on this topic afterwards.

The interest of the European-style options is that they are a class
of derivatives whose Greeks can be computed in closed form for
particular classes of payoff functions. The reason, as we will
show, is that we explicitly know the probability density
function of the random variable involved, $S_T$, whereas in
other scenarios this is not true. This
peculiarity provides us with a framework where we can easily test
how Malliavin Calculus applies to the computation of Greeks. Later,
we will also make a comment on a case where this closed formulas are not available
and where this technique may prove useful.
\subsection{The Malliavin expressions}
Let us start deriving the formal expressions for some of the
Greeks we shall deal with. 

First we assume that our underlying asset $S$ is described by a geometric Brownian motion under the risk neutral probability $\mathbf{P}$:
\begin{equation}
S_{t}=S_{0}+\int_{0}^{t}rS_{s}ds+\int_{0}^{t}\sigma S_{s}\;dW_{s},\label{SI}
\end{equation} where $r$ is the interest rate and
$\sigma$ is the volatility. This model is one of the models
typically used to describe stock prices or stock indices.

Second, from the previous arguments it follows that $X(\alpha)$ must be in general a functional of $S$. In the case of European-type options, $X(\alpha)=S_T$ and from (\ref{SI}):
\begin{equation}
S_T=S_0 e^{\{\mu T + \sigma W_T\}}, \label{ST}
\end{equation}
where $\{W_t\}_{t\in[0,T]}$ is the Wiener process, and $\mu$ is just $r -\sigma^2/2$. Expression (\ref{ST}) is involved in all the following derivations. 

Now we can compute {\it Delta\/}, $\Delta$, the first partial
derivative of the (discounted) expected outcome of the option,
with respect to the present value of the asset:
\begin{equation*}
\Delta=\frac{\partial}{\partial S_0} E \left[e^{-r T}\Phi(S_T)\right] = \frac{e^{-r T}}{S_0} E\left[\frac{\partial S_T}{\partial S_0}\Phi^{\prime}(S_T)\right]=\frac{e^{-r T}}{S_0} E \left[\Phi^{\prime}(S_T) S_T\right].
\end{equation*}
Now we may perform the integration by parts applying the formula
given in (\ref{D1}),
\begin{equation}
\Delta= \frac{e^{-r T}}{S_0} E\left[\Phi(S_T)D^{*}\left(\frac{S_T}{\int_{0}^{T} D_{v} S_T d v}\right)\right], \label{Delta_E_1}
\end{equation}
which removes the derivative of $\Phi$ from the expectation.

The integral term appearing in the last expression will appear many times along our exposition. In order to compute it we must remember the rules
 of the stochastic derivative stated above:
\begin{equation*}
D_{u} S_T = \sigma S_T D_u W_T =  \sigma S_T {\bf 1}_{u\leq T},
\end{equation*}
and then
\begin{equation}
\int_0^T D_{u} S_T = \sigma T S_T. \label{Int_1}
\end{equation}
Then we are able to perform the stochastic integral in
(\ref{Delta_E_1}),
\begin{equation*}
D^{*}\left(\frac{S_T}{\int_{0}^{T} D_{v} S_T d v}\right)=D^{*}\left(\frac{S_T}{\int_{0}^{T} \sigma S_T d v}\right)=D^{*}\left(\frac{1}{\sigma T}\right) = \frac{W_T}{\sigma T},
\end{equation*}
with the help of equation (\ref{D2}) applied to $F=\frac{1}{\sigma T}$. Then the expression for $\Delta$
reads,
\begin{equation}
\Delta=E\left[e^{-rT} \Phi(S_T) \frac{W_{T}}{S_{0}\sigma T}\right].
\end{equation}

Let us move now into a new Greek: {\it Vega\/}. It measures how sensitive is the price of the option when the volatility changes,
\begin {equation*}
{\cal V}=\frac{\partial}{\partial \sigma} E[e^{-rT} \Phi(S_T)] = e^{-rT} E\left[\frac{\partial S_T}{\partial \sigma} \Phi^{\prime}(S_T)\right]=e^{-rT} E[S_T(W_T-\sigma T) \Phi^{\prime}(S_T)].
\end{equation*}
We invoke again the recipe in Section 2 and thus we can
withdraw the derivative form $\Phi(S_T)$,
\begin{equation*}
{\cal V}=e^{-rT} E\left[\Phi(S_T)D^{*}\left(\frac{S_T(W_T-\sigma T)}{\int_{0}^{T} D_{v} S_T d v}\right)\right]=e^{-rT} E\left[\Phi(S_T)D^{*}\left(\frac{W_T}{\sigma T}-1\right)\right],
\end{equation*}
where we have used the expression (\ref{Int_1}). So the
computation we must face is
\begin{equation*}
D^{*}\left(\frac{W_T}{\sigma T}-1\right)=\frac{1}{\sigma T}D^{*}(W_T)-W_T.
\end{equation*}
Here a new instance of stochastic integral appears, $D^{*}(W_T)$.
The rule which we must take into account in order to solve the
problem is again in (\ref{D2}), with $F=W_T$,
\begin{equation*}
D^{*}(W_T)=W_T^2-\int_0^T D_s W_T ds=W_T^2-T,
\end{equation*}
what lead us finally to this expression for ${\cal V}$,
\begin{equation}
{\cal V}=E\left[e^{-rT} \left\{\frac{W_T^2}{\sigma T}-W_T-\frac{1}{\sigma}\right\} \Phi(S_T)\right]. \label{V_E_1}
\end{equation}
The last example we will present here is one involving a second derivative: {\it Gamma\/}. $\Gamma$ inform us on the second order dependence of the price of the option on the actual value of the underlying,
\begin{equation*}
\Gamma=\frac{\partial^2}{\partial S_0^2}E[e^{-rT} \Phi(S_T)]=\frac{e^{-rT}}{S_0^2}E[ S_T^2  \Phi^{\prime\prime}(S_T)].
\end{equation*}
After a first integration by parts we obtain,
\begin{equation*}
\Gamma=\frac{e^{-rT}}{S_0^2}E\left[\Phi^{\prime}(S_T)D^{*}\left(\frac{S_T^2}{\int_{0}^{T} D_{v} S_T d v}\right)\right]=\frac{e^{-rT}}{S_0^2}E\left[\Phi^{\prime}(S_T)D^{*}\left(\frac{S_T}{\sigma T}\right)\right].
\end{equation*}
The stochastic integral
 may be simplified using once more formula (\ref{D2}) on $F=\frac{S_T}{\sigma T}$, leading to 
\begin{equation*}
D^{*}\left(\frac{S_T}{\sigma T}\right)=\frac{S_T}{\sigma T}D^{*}(1) - \frac{1}{\sigma T} \int_0^T D_s S_T ds=S_T \left\{\frac{W_T}{\sigma T} -1\right\}.
\end{equation*}
Afterwards we can perform the second integration by parts,
yielding:
\begin{equation*}
\Gamma=\frac{e^{-rT}}{S_0^2}E\left[\Phi^{\prime}(S_T)S_T \left\{\frac{W_T}{\sigma T} -1\right\}\right]=\frac{e^{-rT}}{S_0^2}E\left[\Phi(S_T)D^{*}\left(\frac{S_T}{\int_{0}^{T} D_v S_T d v} \left\{\frac{W_T}{\sigma T} -1\right\}\right)\right].
\end{equation*}
The stochastic integral is now slightly cumbersome, but it does
not endow any complexity that we have not seen before,
\begin{equation*}
D^{*}\left(\frac{S_T}{\int_{0}^{T} D_v S_T d
v}\left\{\frac{W_T}{\sigma T} -1\right\}\right)=\frac{1}{\sigma
T} D^{*}\left(\frac{W_T}{\sigma T} -1\right)=\frac{1}{\sigma T}
\left\{ \frac{W_T}{\sigma T} - W_T - \frac{1}{\sigma} \right\}.
\end{equation*}
If we bring together the previous partial results we will obtain the expression,
\begin{equation}
\Gamma=E\left[\frac{e^{-rT}}{S_0^2 \sigma T}\left\{ \frac{W_T}{\sigma T} - W_T - \frac{1}{\sigma} \right\}\Phi(S_T) \right]. \label{G_E_1}
\end{equation}
If we compare it with
(\ref{V_E_1}), we find the following relationship between ${\cal V}$ and
$\Gamma$:
\begin{equation}
{\Gamma}=\frac{\cal V}{S_0^2 \sigma T}. \label{V-G}
\end{equation}
Since we have indeed closed expressions for all the
Greeks, we may easily check the correctness of the above
statements. We shall recover
property (\ref{V-G}) of the European-style options in the next section. The
above identities are very well known by practitioners although
their proofs do not usually recall the integration by parts
formula in the form we have introduced it here.

\subsection{The explicit computation}
The reason for the existence of such expressions for the
Greeks of European-type options is due to the fact that there is a
closed and tractable formula for the probability density function
of $S_T$. This is the lognormal distribution that is written as
\begin{equation*}
p(x)=\frac{1}{x \sqrt{2 \pi \sigma^2 T}} \exp\{-[\log(x/S_0)-\mu T]^2 / 2\sigma^2 T\}.
\end{equation*}
When $p(x)$ is available we can face the problem from a different perspective. In this case we are able to compute all the partial derivatives, starting from the explicit formulation for the price of the option, ${\cal P}$,
\begin{equation}
{\cal P}=E[e^{-rT} \Phi(S_T)]= \int_0^{\infty} e^{-rT} \Phi(x) p(x) dx,
\end{equation}
usually just a formal expression, which now becomes handy. We can show this computing the value of $\Delta$, in terms of partial derivatives of $p(x)$:
\begin{equation*}
\Delta = \frac{\partial}{\partial S_0} \int_0^{\infty} e^{-rT} \Phi(x) p(x) dx= \int_0^{\infty}  e^{-rT} \Phi(x) \frac{\partial p(x)}{\partial S_0} dx= \int_0^{\infty} e^{-rT} \Phi(x) \frac{\partial \log p(x)}{\partial S_0} p(x) dx.
\end{equation*}
Note that we get an expression that can be easily rewritten in a
way that apparently resembles our previous results, since one
integration by parts has been implicitly done, and a kernel
naturally appears,
\begin{equation}
\Delta=E\left[e^{-rT} \Phi(S_T) \left(\frac{\partial \log p(x)}{\partial S_0}\right)_{x=S_T}\right]. \label{D_p}
\end{equation}
But we have not yet exploited the information we have about the functional form of $p(x)$,
\begin{equation*}
\left(\frac{\partial \log p(x)}{\partial S_0}\right)_{x=S_T}=\frac{1}{S_0 \sigma^2 T} \left[\log(x/S_0)-\mu T \right]_{x=S_T}=\frac{W_T}{S_0 \sigma T},
\end{equation*}
which leads us to the same expression we have already obtain by
means of Malliavin Calculus:
\begin{equation*}
\Delta=E\left[e^{-rT} \Phi(S_T) \frac{W_{T}}{S_{0}\sigma T}\right].
\end{equation*}

A similar procedure applies to the other Greeks. We will obtain
{\it Vega\/} just replacing the $S_0$ with a $\sigma$ in equation
(\ref{D_p}),
\begin{equation*}
{\cal V}
=E\left[e^{-rT} \Phi(S_T) \left(\frac{\partial \log p(x)}{\partial \sigma}\right)_{x=S_T}\right],
\end{equation*}
and, after straightforward computations, we recover
equation~(\ref{V_E_1}). The case of {\it Gamma\/} leads to an
expression with a very similar flavor to what we have already seen,
\begin{equation*}
\Gamma =E\left[e^{-rT} \Phi(S_T) \left\{\left(\frac{\partial \log p(x)}{\partial S_0}\right)^2+\frac{\partial^2 \log p(x)}{\partial S_0^2} \right\}_{x=S_T}\right],
\end{equation*}
that yields, again, the same result presented in (\ref{G_E_1}). We find therefore in this frame that the property stated in (\ref{V-G}) is fulfilled by {\it Vega\/} and {\it Gamma\/}.

We can then conclude that when we deal with European-style options, the Malliavin-related procedures presented above are equivalent to the result we attain if we directly differenciate the probability density function.

\subsection{The {\em vanilla\/} options} 
\begin{figure}[htb]
\begin{center}
\includegraphics[width=12cm,keepaspectratio=true]{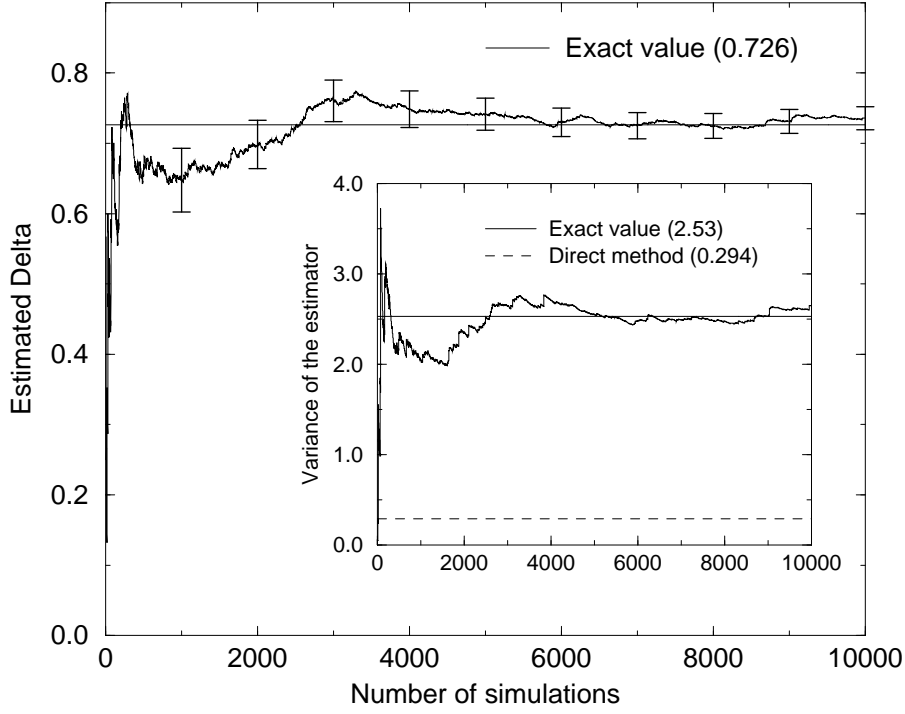}
\caption{Estimated value of {\it Delta\/} for an European call with parameters
 $r=0.1$, $\sigma=0.2$, $T=1.0$ (in years) and $S_0=K=100$
 (in arbitrary cash units), using Monte Carlo techniques.
 } \label{Fig_E_D}
\end{center}
\end{figure}

\begin{figure}[htb]
\begin{center}
\includegraphics[width=12cm,keepaspectratio=true]{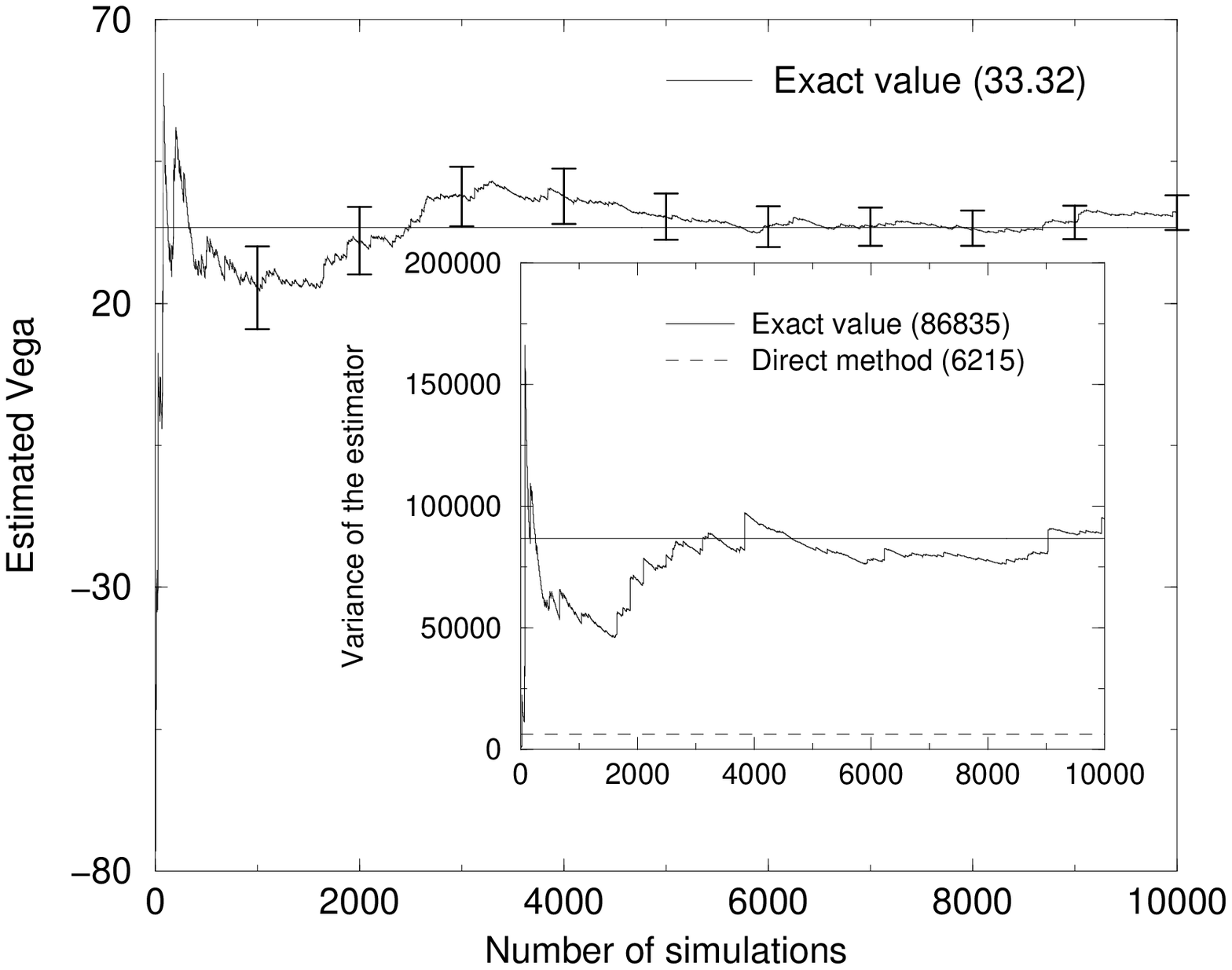}
\caption{Estimated value of {\it Vega\/} for an European call with parameters
 $r=0.1$, $\sigma=0.2$, $T=1.0$ (in years) and $S_0=K=100$
 (in arbitrary cash units), using Monte Carlo techniques.} \label{Fig_E_V}
\end{center}
\end{figure}

Besides the formal comparison with the previous case, the fact of knowing $p(x)$
allows us, in principle, to completely compute all the Greeks once
a payoff function has been selected.
One of the most popular choice is the European, or {\it vanilla\/}, call whose payoff reads,
\begin{equation}
\Phi(X)=(X-K)_{+}. \label{payoff}
\end{equation}
Then can be easily derived the following expressions for the Greeks we have presented:
\begin{eqnarray*}
\Delta&=&\frac{1}{\sqrt{2 \pi}}\int_{-\infty}^{d_1(K)} e^{-x^2/2} dx, \\
{\cal V}&=&S_0 \sqrt{\frac{T}{2 \pi}} e^{-[d_1(K)]^2/2}, \text{ and}\\
\Gamma&=&\frac{1}{S_0 \sqrt{2 \pi \sigma^2 T}} e^{-[d_1(K)]^2/2};
\end{eqnarray*}
where
\begin{equation*}
d_1(x)=\frac{1}{\sigma \sqrt{T}} \left[\log(S_0/x)+(r+\frac{1}{2}\sigma^2)T\right],
\end{equation*}
as it can be found in any textbook on financial derivatives \cite{Wilmott}. In conclussion, we are able to compute the different Greeks using the Malliavin-related formulas, and compare them with their theoretical values. We present in Fig.~\ref{Fig_E_D}~and~\ref{Fig_E_V} the result of this procedure, for a given set of parameters, after Monte Carlo simulation. Only $\Delta$ and ${\cal V}$ are shown, since $\Gamma$ would just be a replica of the second, due to equation (\ref{V-G}). These examples show us how the outcome of the simulation progressively attains their own theoretical value, whereas the statistical error reduces. We notice however that the use of what we have labeled as ``direct method", just performing Monte Carlo simulations starting from the rhs expression in (\ref{Direct}), would lead to an estimator with smaller variance, and therefore a better estimation. Those estimations do not appear in the Figures, in the sake of clarity. But we must remember that this technique can only be applied when the payoff is smooth enough. In our case, when payoff follows (\ref{payoff}), {\it Gamma\/} cannot be computed in this way. 

\section{The Asian-style options}

Here one considers the Greeks for options written on the average
of the stock price $\frac{1}{T}\int_{0}^{T}S_{s}ds$, instead of the final value $S_T$,  as in European options. Note that in
this particular case the density function of the random variable
does not have a known closed formula. {\it Delta\/} in this case is given
by
\begin{equation*}
\Delta=\frac{\partial}{\partial S_0} E \left[e^{-r T}\Phi\left(\frac{1}{T}\int_{0}^{T}S_{s}ds\right)\right] = \frac{e^{-r T}}{S_0} E\left[\Phi^{\prime}\left(\frac{1}{T}\int_{0}^{T}S_{s}ds\right)\frac{1}{T}\int_{0}^{T}S_{u}du\right].
\end{equation*}
There are various ways of doing the integration by parts. In the already cited literature we find in \cite{Fournie99} the following expression:
\begin{equation*}
\Delta=\frac{e^{-r T}}{S_{0}}E\left[ \Phi\left(\frac{1}{T}\int_{0}^{T}S_{s}ds\right)\left( \frac{%
2\int_{0}^{T}S_{t}dW_{t}}{\sigma \int_{0}^{T}S_{t}dt}+1\right) \right],
\end{equation*}
whereas a close variant of it, which involves (\ref{SI}), can be found in \cite{Fournie01}: 
\begin{equation*}
\Delta=\frac{2 e^{-r T}}{S_{0}\sigma^2}E\left[ \Phi\left(\frac{1}{T}\int_{0}^{T}S_{s}ds\right)\left( \frac{S_{T}-S_0}{\int_{0}^{T}S_{t}dt}-\mu \right) \right].
\end{equation*}
Of course, we may also use the same approach we have present in the previous sections, and obtain a third one:
\begin{equation*}
\Delta=\frac{e^{-r T}}{S_{0}}E\left[ \Phi\left(\frac{1}{T}\int_{0}^{T}S_{s}ds\right)\left( \frac{1}{%
<S>}\left\{ \frac{W_{T}}{\sigma }+\frac{<S^{2}>}{<S>}\right\} -1\right) \right],
\end{equation*}
where 
\begin{eqnarray*}
<S>&=&\frac{\int_{0}^{T}tS_{t}dt}{\int_{0}^{T}S_{v}dv}, \text{ and} \\
<S^{2}>&=&\frac{\int_{0}^{T}t^{2}S_{t}dt}{\int_{0}^{T}S_{v}dv}, 
\end{eqnarray*}
are something similar to a first two moments. 

\begin{figure}[htb]
\begin{center}
\includegraphics[width=12cm,keepaspectratio=true]{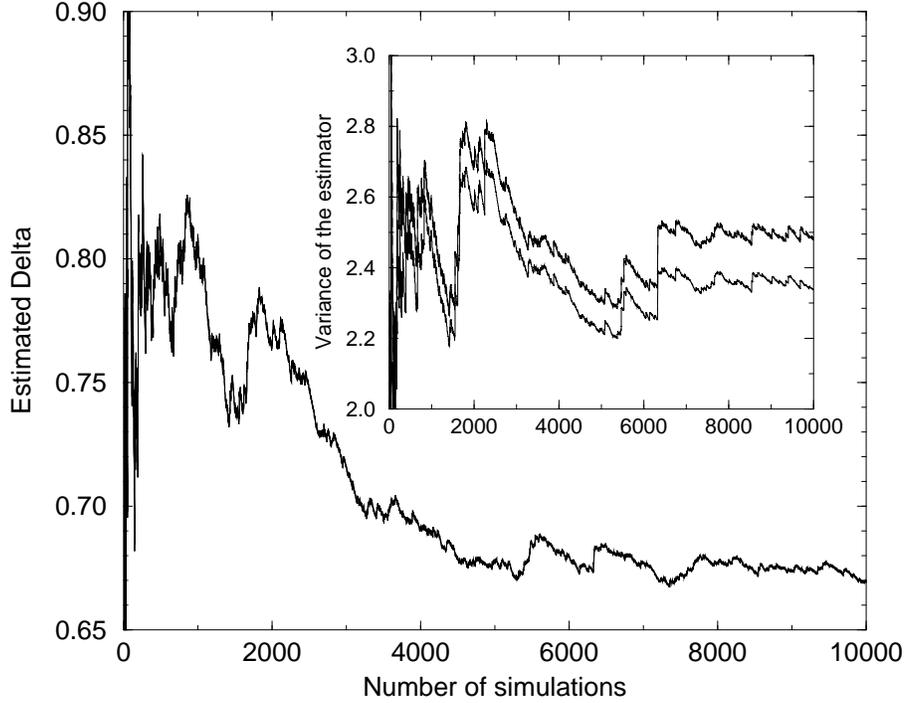}
\caption{Computed value of {\it Delta\/} for an Asian call with parameters
 $r=0.1$, $\sigma=0.2$, $T=1.0$ (in years) and $S_0=K=100$
 (in arbitrary cash units), using Monte Carlo techniques, for the estimators presented in the main text. We have broken the interval of integration in 252 pieces, representing the approximate number of trading days in a year. The exact result turns out to be near the bottom of the graph, at about 0.65.} \label{Fig_A_D}
\end{center}
\end{figure}

Although the two first expressions for $\Delta$ are statistically identical, their particular realizations when perfoming numerical computation will slightly differ, even though the same series of random numbers is used. The last formula is definitely a brand new estimator with its own properties, among them its smaller variance is perhaps the most relevant one.  
We can observe these features in Fig.~\ref{Fig_A_D}, where we show the outcome of the  Monte Carlo simulation using the three alternative instances. We have chosen again the functional form in (\ref{payoff}) for the payoff, and the rest of parameters takes the same value we used in the making of the previous plots.  

Then, not all these formulas coincide and in fact, contrary to what is
claimed in \cite{Fournie01} there is no way to obtain the
integration by parts that provides the minimal variance. The main
reason being that this is equivalent to know the probability
density of the random variable in question. To expose the main
ideas that also appear in \cite{Fournie01} one can note first that
there is an integration by parts that is the ``most''
straightforward but highly unrealistic. For this, consider the
generalized problem
\begin{equation*}
E\left[ \Phi ^{\prime }\left(\int_{0}^{T}S_{s}ds\right)\int_{0}^{T}S_{s}ds\right] =\int_0^{\infty} \Phi ^{\prime }(x)xp(x)dx.
\end{equation*}
Here $p$ denotes the density of $\int_{0}^{T}S_{s}ds$ which exists and is
smooth (it is an interesting exercise of Malliavin Calculus). Therefore one
can perform the integration by parts directly in the above formula thus
obtaining that
\begin{eqnarray*}
E\left[ \Phi ^{\prime }\left(\int_{0}^{T}S_{s}ds\right)\int_{0}^{T}S_{s}ds\right]
&=&\int_0^{\infty} \Phi (x)(p(x)+xp^{\prime }(x))dx \\
&=&E\left[ \Phi \left(\int_{0}^{T}S_{s}ds\right) \cdot \left( 1+\frac{\int_{0}^{T}S_{s}dsp^{%
\prime }(\int_{0}^{T}S_{s}ds)}{p(\int_{0}^{T}S_{s}ds)}\right) \right].
\end{eqnarray*}
Now we procceed to prove that the above gives the minimal integration by
parts in the sense of variance. Obviously it is not possible to carry out
the simulations unless $p^{\prime }$ and $p$ are known. Let us construct the
set of all possible integration by parts. Suppose that $Y$ is a random
variable such that
\begin{equation*}
E\left[ \Phi ^{\prime }\left(\int_{0}^{T}S_{s}ds\right)\int_{0}^{T}S_{s}ds\right] =E%
\left[ \Phi \left(\int_{0}^{T}S_{s}ds\right)Y\right],
\end{equation*}
for any function $\Phi \in C_{p}^{+\infty }$, then it is not difficult to
deduce that
\begin{equation*}
E\left[ Y\left/ \sigma \left( \int_{0}^{T}S_{s}ds\right) \right. \right] =1+%
\frac{\int_{0}^{T}S_{s}dsp^{\prime }(\int_{0}^{T}S_{s}ds)}{%
p(\int_{0}^{T}S_{s}ds)}.
\end{equation*}
Here $\sigma(x)$ denotes the $\sigma$-algebra generated by $x$, and $E[\cdot / \cdot]$ is the conditional expectation. Therefore the set of all possible integration by parts can be characterized
as
\begin{equation*}
\mathcal{M}=\left\{ Y\in L^{2}(\Omega );\text{ }E\left[ Y\left/ \sigma
\left( \int_{0}^{T}S_{s}ds\right) \right. \right] =1+\frac{%
\int_{0}^{T}S_{s}dsp^{\prime }(\int_{0}^{T}S_{s}ds)}{p(\int_{0}^{T}S_{s}ds)}%
\right\}.
\end{equation*}
Next in order we want to find the element in $Y$ that minimizes
\begin{equation*}
\inf_{Y\in \mathcal{M}}E\left[ \Phi \left(\int_{0}^{T}S_{s}ds\right)^{2}Y^{2}\right].
\end{equation*}
As in \cite{Fournie01} is not difficult to see which $Y$ achieves the minimum.
This is done as follows:
\begin{eqnarray*}
E\left[ \Phi \left(\int_{0}^{T}S_{s}ds\right)^{2}Y^{2}\right] &=&E\left[ \Phi
\left(\int_{0}^{T}S_{s}ds\right)^{2}\cdot\left( Y-1-\frac{\int_{0}^{T}S_{s}dsp^{\prime
}(\int_{0}^{T}S_{s}ds)}{p(\int_{0}^{T}S_{s}ds)}\right) ^{2}\right] \\
&&+E\left[ \Phi \left(\int_{0}^{T}S_{s}ds\right)^{2} \cdot \left( 1+\frac{%
\int_{0}^{T}S_{s}dsp^{\prime }(\int_{0}^{T}S_{s}ds)}{p(\int_{0}^{T}S_{s}ds)}%
\right) ^{2}\right],
\end{eqnarray*}
since the mixed product is $0$, due to the property of the set $\mathcal{M}$.
Therefore the minimum is achieved at $Y=\left( 1+\frac{%
\int_{0}^{T}S_{s}dsp^{\prime }(\int_{0}^{T}S_{s}ds)}{p(\int_{0}^{T}S_{s}ds)}%
\right) $. This is clearly impossible to write explicitely as $p$ is unknown
in the case of Asian options. Therefore it is still an open problem to
devise good ways to perform an efficient integration by parts so that the
variance is made small rapidly and efficiently.

\section*{Acknowledgements}
AKH would like to thank Prof. S. Ogawa for his kind invitation to visit Japan, and participate in the workshop. He would also like to thank all the comments received about the presentation of this work. MM has been supported in part by Direcci\'on General de Proyectos de Investigaci\'on under contract No.BFM2000-0795, and by Generalitat de Catalunya under contract No.2000 SGR-00023.

\end{document}